\documentclass[prb,showpacs]{revtex4}
\usepackage{graphicx,psfrag,amssymb,amsmath}
\begin{document}
\title{The magnetic field dependence of the threshold electric field in
unconventional charge density waves}
\author{Bal\'azs D\'ora}
\affiliation{Department of Physics, Technical University of Budapest, H-1521 Budapest, Hungary}
\author{Attila Virosztek}
\affiliation{Department of Physics, Technical University of Budapest, H-1521 Budapest, Hungary}
\affiliation{Research Institute for Solid State Physics and Optics, P.O.Box
49, H-1525 Budapest, Hungary}
\author{Kazumi Maki}
\affiliation{Department of Physics and Astronomy, University of Southern California, Los Angeles CA 90089-0484, USA}

\date{\today}

\begin{abstract}
Many experiments suggest that the unidentified low temperature phase
(LTP) of $\alpha-(BEDT-TTF)_2KHg(SCN)_4$ is most likely unconventional charge
density wave (UCDW).
To further this identification we present our theoretical study of the
threshold electric field of UCDW in a magnetic field. 
The magnetic field-temperature phase diagram is
very similar to those in a d-wave superconductor. We find a rather
strong field dependence of the threshold electric field, which should be readily
accessible experimentally.
\end{abstract}


\maketitle
\section{Introduction}
In recent papers two of us have studied the thermodynamics and the optical
conductivity of unconventional spin density wave (USDW) and unconventional
charge density wave (UCDW) \cite{kiscikk,nagycikk}. Unlike conventional SDW or CDW, we define USDW
and UCDW as the SDW and CDW where the order parameter $\Delta(\bf k)$
depends on the quasi-particle momentum $\bf k$. As to the initial
motivation, the presence of many unconventional superconductors is an
adequate justification \cite{szupravezetes,revmod}. Indeed the thermodynamics of USDW or UCDW
are identical to the one of d-wave superconductor \cite{d-wave}.

On the other hand the low temperature phase (LTP) of $\alpha-(BEDT-TTF)_2KHg(SCN)_4$ has not
been understood until now. The LTP does not exhibit X-ray or NMR signals
characteristic to conventional CDW or SDW \cite{jetp,singl}, and this property is naturally born out
by the UDW model \cite{kiscikk,nagycikk}. The response
of the LTP in a magnetic field suggests that it is not SDW but more likely
a kind of CDW \cite{montambo1,montambo2}. Indeed the phase diagram of the LTP in a magnetic field is
rather similar to the one of a d-wave superconductor in the presence of the
Pauli paramagnetism \cite{phase}.

In the presence of magnetic field (say parallel to the conduction
chain in order to avoid the orbital effect) the LTP splits into two regimes:
the low field regime, where $\Delta(\bf k,r)$ is constant spatially and the
high field regime where $\Delta(\bf k,r)$ varies periodically in space \cite{fflomaki,won1,won2}. In
superconductors the latter regime is called Fulde-Ferrell-Larkin-Ovchinikov
(FFLO) state \cite{fflo1,fflo2}.

Very recently the threshold electric field associated with the sliding
motion of CDW or SDW of $\alpha-(BEDT-TTF)_2KHg(SCN)_4$ has been
reported \cite{ltp}. The temperature dependence of the threshold electric field is
very different from the ones we know for conventional CDW like $NbSe_3$ and
conventional SDW like $(TMTSF)_2PF_6$, though somewhat closer to the one in
SDW \cite{tmtsf}.

We have shown recently \cite{kuszobter} that UDW (either charge or spin) gives the
temperature dependent threshold electric field, which is very close to the
observation. Though the agreement is not perfect, we believe that the
discrepancy is due to imperfect nesting what we neglected for simplicity.

The object of the present paper is to extend the earlier analysis in the
presence of a magnetic field. As in earlier works \cite{montambo1,montambo2,fflomaki,won1,won2}on the related subjects
we focus on the Zeeman splitting (or the Pauli paramagnetic effect) due to
the external magnetic field. Also for simplicity we limit ourselves to the
regime where $\Delta(\bf k,r)$ is spatially uniform. Then the related
thermodynamic quantities have been worked out already in \cite{won1,won2}. Therefore the
model used in \cite{kuszobter}  can be readily extended. We find that the threshold
electric field depends on the external magnetic field as well as on the
temperature. 

Therefore we believe that the predicted field dependence should be readily
accessible experimentally. The $\alpha-(ET)_2$ salts can be put into two
groups: one superconducting and another with this mysterious LTP.
$\alpha-(ET)_2MHg(SCN)_4$ with $M=K$, $Tl$ and $Rb$ belong to the second
group. Hence we suggest that our theory applies to the LTP in the second
group.

\section{Phase diagram and density of states}

\begin{figure}[h!]
\hspace*{2cm}
\psfrag{x}[t][b][1][0]{$T/T_{c0}$}
\psfrag{y}[t][b][1][0]{$h/\Delta_{00}$}
\psfrag{z}[b][t][1][0]{$\Delta(T,h)/\Delta_{00}$}
\vspace*{0cm}
{\includegraphics[width=13cm,height=8cm]{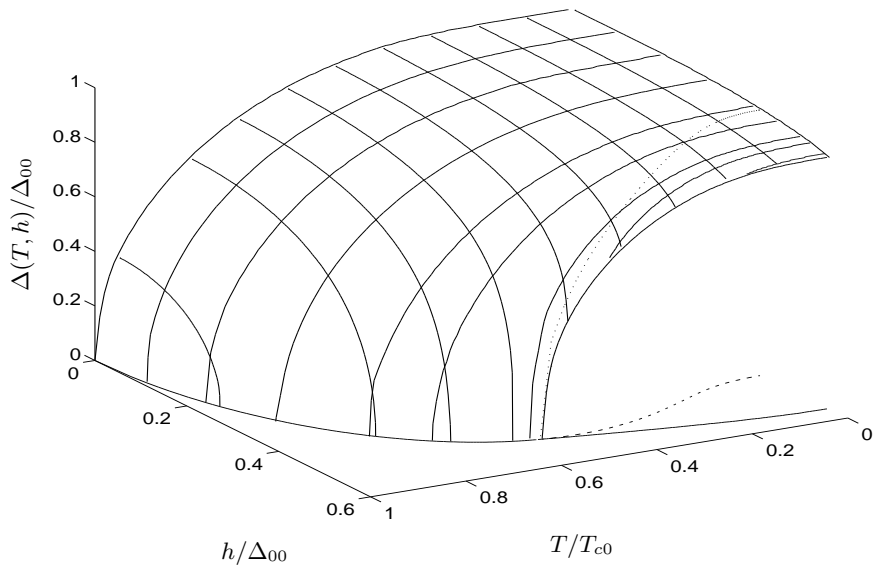}}

\caption{Stereograph of the order parameter in the reduced 
temperature and field plane. The dotted line denotes the metastability line above which
the normal state becomes local minimum of the free energy.\label{fig:delta}}
\end{figure}

The phase diagram is the same as the one in a d-wave superconductor
\cite{phase} without the FFLO state.
At $T=0$ a first order transition occurs to the normal state at $h=0.56\Delta_{00}$, where $\Delta_{00}$ is the zero field zero temperature
order parameter and $h=\mu_B H$. The value of the gap is $0.92\Delta_{00}$
at the transition point. With decreasing
field, the transition occurs at $h=0.41\Delta_{00}$, and the gap
jumps from zero to $0.97\Delta_{00}$. For $T<0.56T_{c0}$ ($T_{c0}$ is the transition temperature at $h=0$) the transition remains first order, and
hysteresis is observable somewhere between $0.41<h/\Delta_{00}<0.56$.
In this region, the normal state becomes local minimum of the free energy,
and depending on the direction of the change of the external field, the first order
transition occurs at smaller field approaching from the normal state than
quitting the DW phase with increasing field.
By exceeding $T\simeq 0.56T_{c0}$, the transition becomes second order at the
bicritical point ($h\approx0.51\Delta_{00}$, $T\approx0.56T_{c0}$).
It is worth noting that the phase diagram is modified at $T<0.56T_{c0}$ and
$h\sim 0.51\Delta_{00}$ because of the possibility of the FFLO regime what we
excluded here for simplicity.
The order parameter as a function of $T$ and $h$ is shown in
Fig. \ref{fig:delta}.

\begin{figure}[h!]
\hspace*{2cm}
\psfrag{x}[t][b][1][0]{$E/\Delta$}
\psfrag{y}[b][t][1][0]{$N(E)/\rho_0$}
\vspace*{0cm}
{\includegraphics[width=13cm,height=8cm]{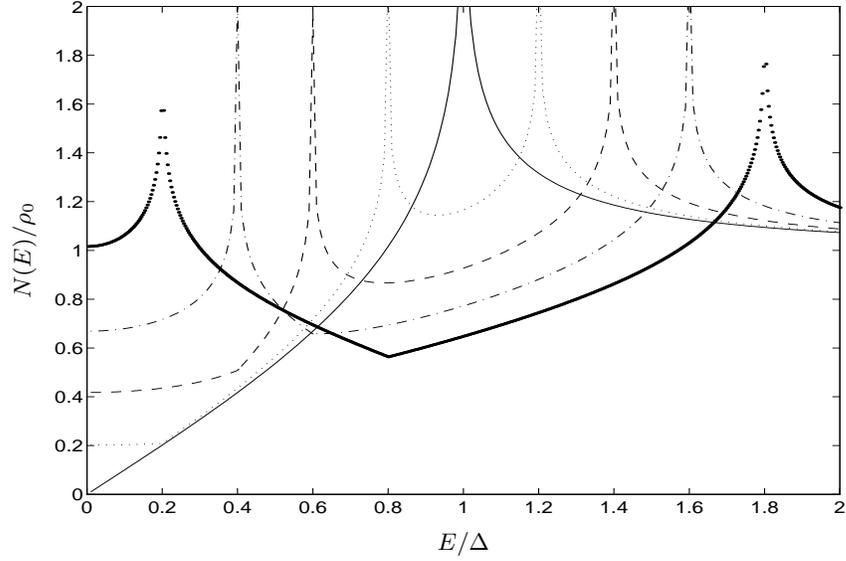}}

\caption{The density of states is shown as a function of $E/\Delta$ for $h/\Delta=0$ (solid line), $0.2$ (thin dotted line), $0.4$ (dashed line), $0.6$ (dashed-dotted line) and $0.8$ (thick dotted line).\label{fig:dosh}}
\end{figure}

The quasiparticle density of states is obtained as \cite{won1}
\begin{equation}
N(E)=\frac 12 (\rho(E+h)+\rho(E-h)),
\end{equation}
where $\rho(E)$ is the density of states in the absence of magnetic field, and is given by\cite{nagycikk} 
$\rho(E)/\rho_0(0)=(2|E|/\pi|\Delta|)K(|E|/|\Delta|)$ if $|E|<|\Delta|$,
and $\rho(E)/\rho_0(0)=(2/\pi)K(|\Delta|/|E|)$ if $|E|>|\Delta|$. $K(z)$ is the complete elliptic integral of the first kind. As $h$ increases, the valley at the Fermi surface is filled in. Also the divergent peaks at $\pm\Delta$ split into four new peaks at $\pm\Delta\pm h$. Interestingly, at $h=\Delta$ the density of states is divergent at the Fermi surface. These properties can be seen in Fig. \ref{fig:dosh}.
  
\section{Phase Hamiltonian and the threshold electric field}

\begin{figure}[h!]
\hspace*{2cm}
\psfrag{x}[t][b][1][0]{$T/T_{c0}$}
\psfrag{y}[t][b][1][0]{$h/\Delta_{00}$}
\psfrag{z}[b][t][1][0]{$\rho_s(T,h)/\rho_s(0,0)$}
\vspace*{0cm}
{\includegraphics[width=13cm,height=8cm]{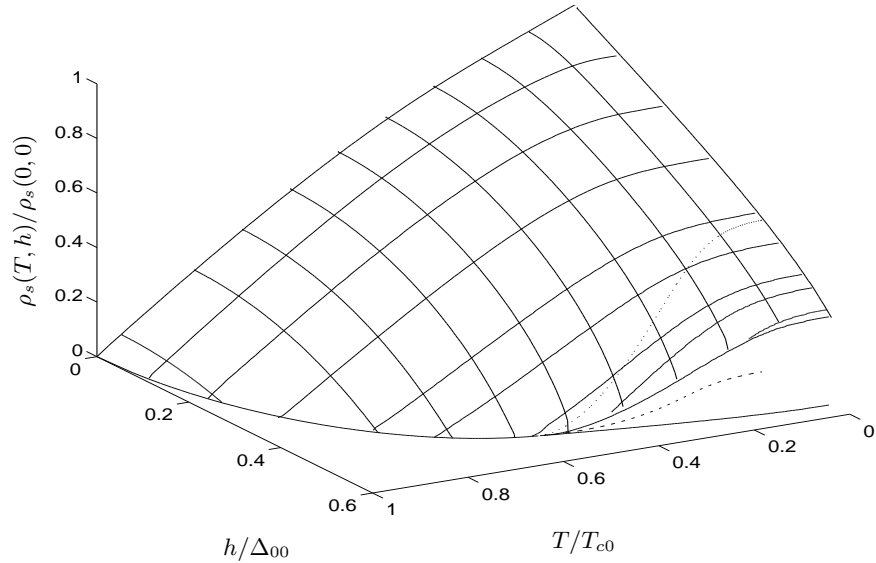}}

\caption{Stereograph of the condensate density in the reduced 
temperature and field plane. The dotted line denotes the metastability line above which
the normal state becomes local minimum of the free energy.\label{fig:sfd3d}}
\end{figure}

It is the most convenient to formulate the threshold electric field in
terms of the phase Hamiltonian which is given as \cite{viro1,viro2}
\begin{eqnarray}
H(\Phi)={\bf \int}d^3r\left\{\frac 1 4 N_0 f \left[ v_F^2 \left(\frac{\partial\Phi} {\partial x}\right)^2
+v_b^2 \left(\frac{\partial\Phi} {\partial y}\right)^2
+v_c^2 \left(\frac{\partial\Phi} {\partial
z}\right)^2+
\left(\frac{\partial\Phi}
{\partial t}\right)^2-4v_FeE\Phi\right]+V_{imp}(\Phi)\right\} \label{phaseH}
\end{eqnarray}
where $N_0$ is
the density of states in the normal state at the Fermi surface per spin,
$f=\rho_s(T,h)/\rho_s(0,0)$ where $\rho_s(T,h)$ is the condensate density
\cite{won1} and $E$
is an electric field applied in the $x$ direction. Here $v_F$, $v_b$ and
$v_c$ are the characteristic velocities of the quasi-one dimensional
electron system in the three spatial directions. For UDW the condensate
density is the same as the superfluid density in d-wave superconductors and is 
shown in Fig. \ref{fig:sfd3d}.

We may think of Eq.(\ref{phaseH}) as a natural extension of the
Fukuyama-Lee-Rice Hamiltonian \cite{fl,lr} for UCDW and for $T\neq0$, $H\neq0$ and for
three spatial directions.

The pinning potential is obtained as
\begin{gather}
V_{imp}(\Phi)=- \frac{8 V_0 V_y N_0^2}{\pi}\sum_j \cos(2({\bf
QR}_j+\Phi({\bf R}_j))) 
\Delta(T,h)\times \nonumber \\
\times \int_0^1\frac12\left(\tanh\frac{\beta(\Delta(T,h)
x+h)}{2}+\tanh\frac{\beta(\Delta(T,h)x-h)}{2}\right) E(\sqrt{1-x^2})
(K(x)-E(x))dx, \label{pinning}
\end{gather}
where ${\bf R}_j$
 is an impurity site, $K(z)$ and $E(z)$ are the complete elliptic
 integrals of the first and second kind, respectively. In obtaining
 Eq. \ref{pinning} we assumed a nonlocal impurity potential \cite{kuszobter}:
\begin{equation}
U({\bf Q+q})=V_0+\sum_{i=y,z}V_i\cos(q_i \delta_i).
\end{equation} 
Then following FLR \cite{fl,lr}, in the strong pinning limit the threshold
electric field at $T=0$ is given by
\begin{equation}
E_T^S(0,h)=\frac{2k_F}{e}\frac{n_i}{n} N_0^2V_0V_y\frac{16}{\pi}
\frac{\Delta(0,h)}{\rho_s(0,h)}\int\limits_{h/\Delta_{00}}^1 E(\sqrt{1-x^2})(K(x)-E(x))dx
\end{equation}
and for general temperature it is obtained as
\begin{gather}
\frac{E_T^S(T,h)}{E_T^S(0,0)}=\frac{\rho_s(0,0)}{\rho_s(T,h)}\frac{\Delta(T,h)}
{\Delta_{00}}
\frac{1}{0.5925} \times \nonumber \\
\times\int_0^1\frac12\left(\tanh\frac{\beta(\Delta(T,h)x+h)}{2}
+\tanh\frac{\beta(\Delta(T,h)x-h)}{2}\right)E(\sqrt{1-x^2})
(K(x)-E(x))dx. 
\end{gather}

\begin{figure}[h!]
\hspace*{2cm}
\psfrag{x}[t][b][1][0]{$T/T_{c0}$}
\psfrag{y}[t][b][1][0]{$h/\Delta_{00}$}
\psfrag{z}[b][t][1][0]{$E_T^S(T,h)/E_T^S(0,0)$}
\vspace*{0cm}
{\includegraphics[width=13cm,height=8cm]{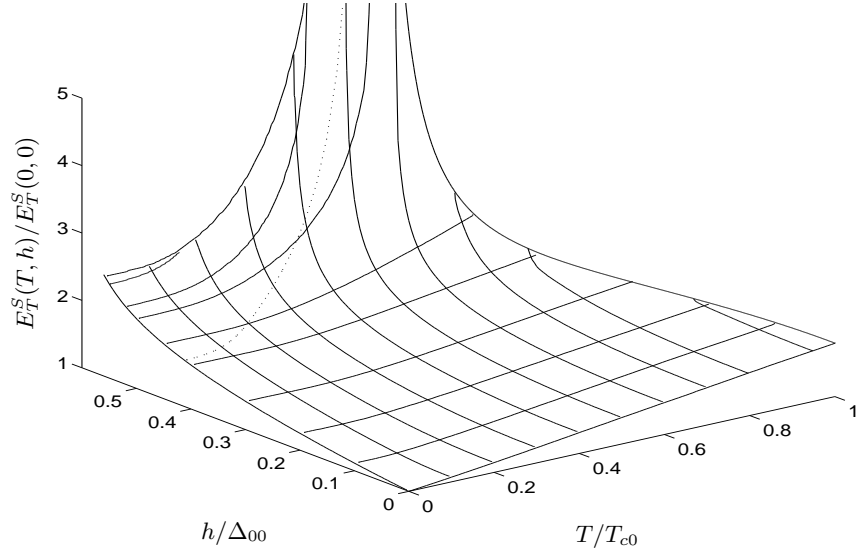}}

\caption{The threshold electric field in the strong pinning limit is plotted as a function of the reduced 
temperature and field. The dashed line is the threshold field belonging to
the metastability line.\label{fig:etvsth}}
\end{figure}

At low temperature $E_T^S$ is well approximated by
\begin{gather}
\frac{E_T^S(T,h)}{E_T^S(0,0)}\approx\frac{\rho_s(0,0)}{\rho_s(T,h)}=\left\{
\begin{array}{lc}
1+2\ln(2) \frac{T}{\Delta_{00}}+\frac{h^2}{4T\Delta_{00}} & \frac hT\ll 1\\
1+\frac{h}{\Delta_{00}} & \frac hT\gg1
\end{array}\right.
\end{gather}
in the $h,T\ll \Delta_{00}$ range \cite{won1}. For $T>0.56T_{c0}$, along the second order phase boundary the threshold
field reads as
\begin{equation}
\frac{E_T^S(T,h)}{E_T^S(0,0)}=-\frac{\textmd{Re}\Psi^\prime\left(\frac12+\frac{ih}{2\pi
T}\right)}{\textmd{Re}\Psi^{\prime\prime}\left(\frac 12+\frac{ih}{2\pi
T}\right)}\frac{T}{\Delta_{00}}\frac{\pi^3}{4\times0.5925},
\end{equation}
which is divergent at the bicritical point (possibly tricritical with the FFLO
state) due to the zero of $\textmd{Re}\Psi^{\prime\prime}\left(\frac 12+
\frac{ih}{2\pi T}\right)$ at $h/T\approx 1.91$. As a result the threshold
electric field
close to the bicritical point is given by
\begin{equation}
\frac{E_T^S(T,h)}{E_T^S(0,0)}=\frac{1.12}{1.91-h/t}
\end{equation}
from the second order phase boundary.
We show the threshold electric field as a function of the temperature and the magnetic field
in Fig. \ref{fig:etvsth} in the strong pinning limit.

\begin{figure}[h!]
\hspace*{2cm}
\psfrag{x}[t][b][1][0]{$T/T_{c0}$}
\psfrag{y}[b][t][1][0]{$E_T^W(T,h)/E_T^W(0,h)$}
\vspace*{0cm}
{\includegraphics[width=13cm,height=8cm]{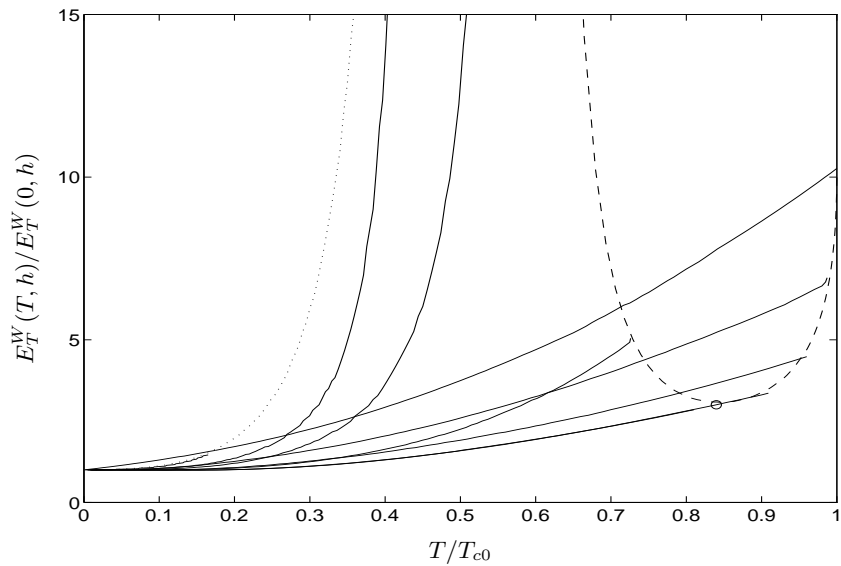}}

\caption{The threshold electric field in the weak pinning limit is plotted as a function of the reduced 
temperature for $h/\Delta_{00}=0$, $0.1$, $0.2$, $0.3$, $0.4$, $0.45$, $0.5$, $0.52$,
$0.55$ with endpoints from right to left. The circle represents the end of the $h=0.4\Delta_{00}$ curve,
which is very close to the $h=0.3\Delta_{00}$ one. The dashed line accounts
for the threshold field along the second order phase boundary while the
dotted for the one along the first order phase boundary.\label{fig:etvst}}
\end{figure}

\begin{figure}[h!]
\hspace*{2cm}
\psfrag{x}[t][b][1][0]{$h/\Delta_{00}$}
\psfrag{y}[b][t][1][0]{$E_T^W(T,h)/E_T^W(T,0)$}
\vspace*{0cm}
{\includegraphics[width=13cm,height=8cm]{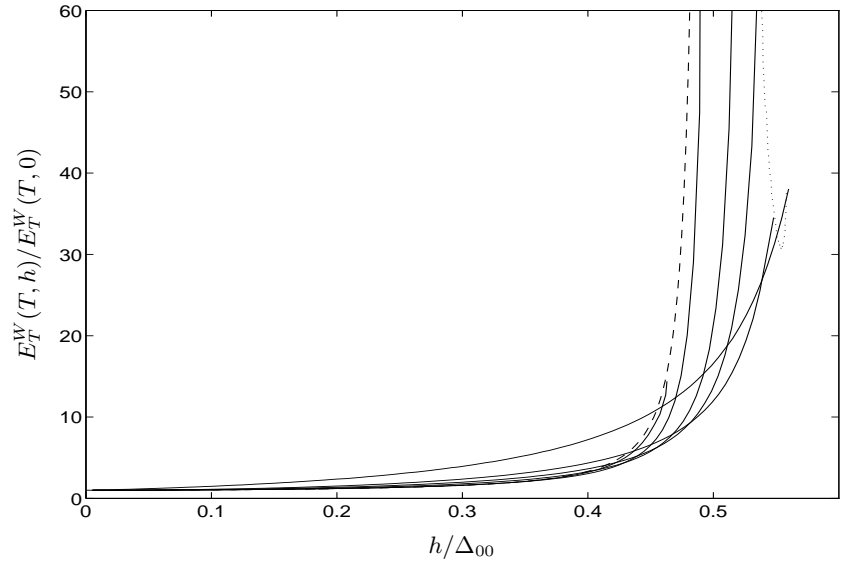}}

\caption{The threshold electric field in the weak pinning limit is plotted
as a function of the magnetic field for $T/T_{c0}=0$, $0.2$, $0.3$, $0.4$,
$0.6$ and $0.7$ with endpoints from right to left. The dashed line accounts
for the threshold field along the second order phase boundary while the
dotted for the one along the first order phase boundary.\label{fig:etvsh}}
\end{figure}

The weak-pinning limit is more appropriate for high quality crystals. Then
 we obtain for a three dimensional system
\begin{equation}
\frac{E_T^W(T,h)}{E_T^W(0,0)}=\left(\frac{E_T^S(T,h)}{E_T^S(0,0)}\right)^4.
\end{equation}
$E_T^W(T,h)$ is plotted as a function of temperature and magnetic field in Figs. \ref{fig:etvst} and \ref{fig:etvsh}, respectively.  
 
At small but increasing fields, the enhancement of the threshold electric
field at the transition
temperature relative to the $T=0$ value becomes smaller due to the initial linear decrease of the condensate density versus $h$ at $T=0$.

At low temperature $E_T^W(T,h)$ increases with $h$ almost linearly. Further
$E_T^W(T,h)$ diverges for $T\simeq0.56 T_{c0}$ when the magnetic transition
changes from second order to first order. Indeed
the strong $H$ dependence of the threshold electric field at $T=2.2$K
is described in \cite{ltp} which appears to be consistent with the present
result,though no detail is available. Unfortunately the present result
does not apply for $T<0.56T_{c0}$ and $h\gtrsim 0.51\Delta_{00}$ due to the
presence of the FFLO regime.
Nevertheless the present result can be tested in a wide range of the $H-T$
phase diagram of the LTP in $\alpha-(ET)_2$ salts. The effect of the FFLO
state and the related threshold electric field will be discussed elsewhere.

\section{Concluding remarks}

We have extended  our earlier analysis of the threshold electric field in
UCDW in the presence of magnetic field. The magnetic field is introduced as
the Zeeman splitting. The phase diagram is found to be identical to the one
in a d-wave superconductor \cite{phase} without the FFLO state. The present
model predicts very strong $H$ dependence of the threshold electric field, even divergent behaviour at the
bicritical point, which should be readily accessible
experimentally.

If such a strong $H$ dependence of $E_T$ is observed, this will surely
strengthen our proposal that the LTP of $\alpha-(ET)_2$ salts should be
UCDW.

\begin{acknowledgments}
We would like to thank Bojana Korin-Hamzi\'c for useful discussions on
possible $H$ dependence of the threshold electric field.
This work
was supported by the Hungarian National Research Fund under grant numbers
OTKA T032162 and T029877, and by the Ministry of Education under grant number
FKFP 0029/1999.
\end{acknowledgments}

\begin{thebibliography}{10}
\expandafter\ifx\csname bibnamefont\endcsname\relax
  \def\bibnamefont#1{#1}\fi
\expandafter\ifx\csname bibfnamefont\endcsname\relax
  \def\bibfnamefont#1{#1}\fi
\expandafter\ifx\csname url\endcsname\relax
  \def\url#1{\texttt{#1}}\fi
\expandafter\ifx\csname urlprefix\endcsname\relax\def\urlprefix{URL }\fi
\providecommand{\bibinfo}[2]{#2}
\providecommand{\eprint}[2][]{\url{#2}}

\bibitem{kiscikk}
\bibinfo{author}{\bibfnamefont{B.}~\bibnamefont{D{\'o}ra}} \bibnamefont{and}
  \bibinfo{author}{\bibfnamefont{A.}~\bibnamefont{Virosztek}},
  \bibinfo{journal}{J. Phys. IV France} \textbf{\bibinfo{volume}{9}},
  \bibinfo{pages}{Pr10-239} (\bibinfo{year}{1999}).

\bibitem{nagycikk}
\bibinfo{author}{\bibfnamefont{B.}~\bibnamefont{D{\'o}ra}} \bibnamefont{and}
  \bibinfo{author}{\bibfnamefont{A.}~\bibnamefont{Virosztek}},
  \bibinfo{note}{cond-mat/0010162, submitted to Eur. Phys. J. B}.

\bibitem{szupravezetes}
\bibinfo{author}{\bibfnamefont{H.}~\bibnamefont{Won}} \bibnamefont{and}
  \bibinfo{author}{\bibfnamefont{K.}~\bibnamefont{Maki}}, in
  \emph{\bibinfo{booktitle}{Symmmetry and Pairing in Superconductors}}, edited
  by \bibinfo{editor}{\bibfnamefont{M.}~\bibnamefont{Ausloos}}
  \bibnamefont{and} \bibinfo{editor}{\bibfnamefont{S.}~\bibnamefont{Kruchinin}}
  (\bibinfo{publisher}{Kluwer}, \bibinfo{address}{Dordrecht},
  \bibinfo{year}{1999}).

\bibitem{revmod}
\bibinfo{author}{\bibfnamefont{C.~C.} \bibnamefont{Tsuei}} \bibnamefont{and}
  \bibinfo{author}{\bibfnamefont{J.~R.} \bibnamefont{Kirtley}},
  \bibinfo{journal}{Rev. Mod. Phys.} \textbf{\bibinfo{volume}{72}},
  \bibinfo{pages}{969} (\bibinfo{year}{2000}).

\bibitem{d-wave}
\bibinfo{author}{\bibfnamefont{H.}~\bibnamefont{Won}} \bibnamefont{and}
  \bibinfo{author}{\bibfnamefont{K.}~\bibnamefont{Maki}},
  \bibinfo{journal}{Phys. Rev. B} \textbf{\bibinfo{volume}{49}},
  \bibinfo{pages}{1397} (\bibinfo{year}{1994}).

\bibitem{jetp}
\bibinfo{author}{\bibfnamefont{P.}~\bibnamefont{Christ}},
  \bibinfo{author}{\bibfnamefont{W.}~\bibnamefont{Biberacher}},
  \bibinfo{author}{\bibfnamefont{M.~V.} \bibnamefont{Kartsovnik}},
  \bibinfo{author}{\bibfnamefont{E.}~\bibnamefont{Steep}},
  \bibinfo{author}{\bibfnamefont{E.}~\bibnamefont{Balthes}},
  \bibinfo{author}{\bibfnamefont{H.}~\bibnamefont{Weiss}}, \bibnamefont{and}
  \bibinfo{author}{\bibfnamefont{H.}~\bibnamefont{M{\"u}ller}},
  \bibinfo{journal}{JETP Lett.} \textbf{\bibinfo{volume}{71}},
  \bibinfo{pages}{303} (\bibinfo{year}{2000}).

\bibitem{singl}
\bibinfo{author}{\bibfnamefont{J.}~\bibnamefont{Singleton}},
  \bibinfo{journal}{Rep. Prog. Phys.} \textbf{\bibinfo{volume}{63}},
  \bibinfo{pages}{1161} (\bibinfo{year}{2000}).

\bibitem{montambo1}
\bibinfo{author}{\bibfnamefont{D.}~\bibnamefont{Zanchi}},
  \bibinfo{author}{\bibfnamefont{A.}~\bibnamefont{Bjelis}}, \bibnamefont{and}
  \bibinfo{author}{\bibfnamefont{G.}~\bibnamefont{Montambaux}},
  \bibinfo{journal}{Phys. Rev. B} \textbf{\bibinfo{volume}{53}},
  \bibinfo{pages}{1240} (\bibinfo{year}{1996}).

\bibitem{montambo2}
\bibinfo{author}{\bibfnamefont{A.}~\bibnamefont{Bjelis}},
  \bibinfo{author}{\bibfnamefont{D.}~\bibnamefont{Zanchi}}, \bibnamefont{and}
  \bibinfo{author}{\bibfnamefont{G.}~\bibnamefont{Montambaux}},
  \bibinfo{journal}{J. Phys. IV France} \textbf{\bibinfo{volume}{9}},
  \bibinfo{pages}{Pr10-203} (\bibinfo{year}{1999}).

\bibitem{phase}
\bibinfo{author}{\bibfnamefont{K.}~\bibnamefont{Yang}} \bibnamefont{and}
  \bibinfo{author}{\bibfnamefont{S.~L.} \bibnamefont{Sondhi}},
  \bibinfo{journal}{Phys. Rev. B} \textbf{\bibinfo{volume}{57}},
  \bibinfo{pages}{8566} (\bibinfo{year}{1998}).

\bibitem{fflomaki}
\bibinfo{author}{\bibfnamefont{K.}~\bibnamefont{Maki}} \bibnamefont{and}
  \bibinfo{author}{\bibfnamefont{H.}~\bibnamefont{Won}},
  \bibinfo{journal}{Czech. J. Phys.} \textbf{\bibinfo{volume}{46}},
  \bibinfo{pages}{S2, 1035} (\bibinfo{year}{1996}).

\bibitem{won1}
\bibinfo{author}{\bibfnamefont{H.}~\bibnamefont{Won}},
  \bibinfo{author}{\bibfnamefont{H.}~\bibnamefont{Jang}}, \bibnamefont{and}
  \bibinfo{author}{\bibfnamefont{K.}~\bibnamefont{Maki}},
  \bibinfo{note}{cond-mat/9901252}.

\bibitem{won2}
\bibinfo{author}{\bibfnamefont{H.}~\bibnamefont{Won}}, ,
  \bibinfo{author}{\bibfnamefont{H.}~\bibnamefont{Jang}}, \bibnamefont{and}
  \bibinfo{author}{\bibfnamefont{K.}~\bibnamefont{Maki}},
  \bibinfo{journal}{Physica B} \textbf{\bibinfo{volume}{281-282}},
  \bibinfo{pages}{944} (\bibinfo{year}{2000}).

\bibitem{fflo1}
\bibinfo{author}{\bibfnamefont{P.}~\bibnamefont{Fulde}} \bibnamefont{and}
  \bibinfo{author}{\bibfnamefont{R.~A.} \bibnamefont{Ferrell}},
  \bibinfo{journal}{Phys. Rev.} \textbf{\bibinfo{volume}{35}},
  \bibinfo{pages}{A550} (\bibinfo{year}{1964}).

\bibitem{fflo2}
\bibinfo{author}{\bibfnamefont{A.~I.} \bibnamefont{Larkin}} \bibnamefont{and}
  \bibinfo{author}{\bibfnamefont{N.}~\bibnamefont{Ovchinikov}},
  \bibinfo{journal}{Sov. Phys. JETP} \textbf{\bibinfo{volume}{20}},
  \bibinfo{pages}{762} (\bibinfo{year}{1965}).

\bibitem{ltp}
\bibinfo{author}{\bibfnamefont{M.}~\bibnamefont{Basleti{\'c}}},
  \bibinfo{author}{\bibfnamefont{B.}~\bibnamefont{Korin-Hamzi{\'c}}},
  \bibinfo{author}{\bibfnamefont{M.~V.} \bibnamefont{Kartsovnik}},
  \bibnamefont{and}
  \bibinfo{author}{\bibfnamefont{H.}~\bibnamefont{M{\"u}ller}},
  \bibinfo{note}{{\MakeUppercase{S}}ynth. Metals, in press}.

\bibitem{tmtsf}
\bibinfo{author}{\bibnamefont{see for~example S.~Tomi{\'c}}},
  \bibinfo{author}{\bibfnamefont{J.~R.} \bibnamefont{Cooper}},
  \bibinfo{author}{\bibfnamefont{W.}~\bibnamefont{Kang}},
  \bibinfo{author}{\bibfnamefont{D.}~\bibnamefont{Jerome}}, \bibnamefont{and}
  \bibinfo{author}{\bibfnamefont{K.}~\bibnamefont{Maki}}, \bibinfo{journal}{J.
  Phys. I (Paris)} \textbf{\bibinfo{volume}{1}}, \bibinfo{pages}{1603}
  (\bibinfo{year}{1991}).

\bibitem{kuszobter}
\bibinfo{author}{\bibfnamefont{B.}~\bibnamefont{D{\'o}ra}},
  \bibinfo{author}{\bibfnamefont{A.}~\bibnamefont{Virosztek}},
  \bibnamefont{and} \bibinfo{author}{\bibfnamefont{K.}~\bibnamefont{Maki}},
  \bibinfo{note}{cond-mat/0101332, submitted to Phys. Rev. B}.

\bibitem{viro1}
\bibinfo{author}{\bibfnamefont{K.}~\bibnamefont{Maki}} \bibnamefont{and}
  \bibinfo{author}{\bibfnamefont{A.}~\bibnamefont{Virosztek}},
  \bibinfo{journal}{Phys. Rev. B} \textbf{\bibinfo{volume}{39}},
  \bibinfo{pages}{9640} (\bibinfo{year}{1989}).

\bibitem{viro2}
\bibinfo{author}{\bibfnamefont{K.}~\bibnamefont{Maki}} \bibnamefont{and}
  \bibinfo{author}{\bibfnamefont{A.}~\bibnamefont{Virosztek}},
  \bibinfo{journal}{Phys. Rev. B} \textbf{\bibinfo{volume}{42}},
  \bibinfo{pages}{655} (\bibinfo{year}{1990}).

\bibitem{fl}
\bibinfo{author}{\bibfnamefont{H.}~\bibnamefont{Fukuyama}} \bibnamefont{and}
  \bibinfo{author}{\bibfnamefont{P.~A.} \bibnamefont{Lee}},
  \bibinfo{journal}{Phys. Rev. B} \textbf{\bibinfo{volume}{17}},
  \bibinfo{pages}{535} (\bibinfo{year}{1978}).

\bibitem{lr}
\bibinfo{author}{\bibfnamefont{P.~A.} \bibnamefont{Lee}} \bibnamefont{and}
  \bibinfo{author}{\bibfnamefont{T.~M.} \bibnamefont{Rice}},
  \bibinfo{journal}{Phys. Rev. B} \textbf{\bibinfo{volume}{19}},
  \bibinfo{pages}{3970} (\bibinfo{year}{1979}).

\end{thebibliography}

\end{document}